%% file: main.tex
\documentclass{article}

\usepackage{arxiv}

\input{macros}

\title{Smart-Corpus: an Organized Repository of Ethereum Smart Contracts Source Code and Metrics}

\newcommand{\bc}{blockchain\xspace}
\newcommand{\ES}{EtherScan\xspace}
\newcommand{\Eth}{Ethereum\xspace}
\newcommand{\SC}{Smart Corpus\xspace}
\newcommand{\sco}{smart-contract\xspace}
\newcommand{\scs}{smart-contracts\xspace}
\newcommand{\Scs}{Smart-contracts\xspace}
\newcommand{\SChtmlGUI}{https://aphd.github.io/smac-corpus/}
\newcommand{\Sol}{Solidity\xspace}

\author{
    Giuseppe Antonio Pierro  \\
    Inria Lille - Nord Europe\\
    Villeneuve D’ascq, France \\
  \texttt{giuseppe.pierro@inria.fr} \\
   \And
   Roberto Tonelli \\
   Department of Mathematics and Computer Science \\
   University of Cagliari\\
   09124 Cagliari, Italy\\
  \texttt{sroberto.tonelli@dsf.unica.it} \\
  \And
  Michele Marchesi \\
  Department of Mathematics and Computer Science \\
  University of Cagliari\\
  09124 Cagliari, Italy\\
 \texttt{marchesi@unica.it} \\
}

\begin{document}
\maketitle

% Abstract (Do not insert blank lines, i.e. \\) 
\begin{abstract}
Many empirical software engineering studies show that there is a great need for repositories where source code is acquired, filtered and classified. During the last few years, Ethereum block explorer services have emerged as a popular project to explore and search Ethereum blockchain data such as transactions, addresses, tokens, smart-contracts’ source code, prices and other activities taking place on the Ethereum blockchain. Despite the availability of this kind of services, retrieving specific information useful to empirical software engineering studies, such as the study of smart-contracts’ software metrics might require many sub-tasks, such as searching specific transactions in a block, parsing files in HTML format and filtering the smart-contracts to remove duplicated code or unused smart-contracts.\\
In this paper we afford this problem creating Smart Corpus’, a Corpus of Smart Contracts in an organized reasoned and up to date repository where Solidity source code and other metadata about Ethereum smart contracts can easily and systematically be retrieved. We present the Smart Corpus’ design and its initial implementation and we show how the data-set of smart contracts’ source code in a variety of programming languages can be queried and processed, get useful information on smart contracts and their software metrics.
The Smart Corpus aims to create a smart-contracts’ repository where smart contracts data (source code, ABI and byte-code) are freely and immediately available and also classified based on the main software metrics identified in the scientific literature. Smart-contracts source code has been validated by EtherScan and each contract comes with its own associated software metrics as computed by the freely available software PASO. Moreover, Smart Corpus can be easily extended, as the number of new smart-contracts increases day by day.
\end{abstract}
% Keywords
% List three to ten pertinent keywords specific to the article, yet reasonably common within the subject discipline.

\section{Introduction}

With the advent of \bc technology as a mainstream technological innovation many researchers and software developers started investigating the new possibilities for software products relying on such infrastructure. Second generation blockchains offer the possibility to code the so called \scs in a Turing complete programming language where all the main operations of traditional software systems can be carried on. 
The paradigmatic reference is the Ethereum \bc which offers the possibility to deploy and execute decentralized applications (dApps) which are mainly coded in Solidity, presently the most adopted programming language \cite{O_Donovan_2019}.

Coding \scs which run in a \bc environment has its peculiarities and constraints and differs from coding in traditional out-of-chain contexts~\cite{pierro20}. One of the major differences is the immutability of deployed code, so that if bugs or bad smells are introduced into a \sco these cannot be fixed afterwards with patches. Another contract must be deployed in substitution of the former and users must be well 
advised not to use the wrong code. Another main issue is the interaction with the \bc by means of transactions where information exchange can occur only between \bc internal components. Furthermore, memory occupation on blockchain typically has a cost that developers want to reduce and 
chaining all the blocks poses limits to the reasonable space available for each \sco imposing practical constraints to source code size~\cite{Murgia2010}. 

This new programming paradigm poses major challenges also to expert developers so that famous failures are commonly found in \bc software~\cite{Zheng_2020, 9127445, Murgia12}. 
The novelty of the paradigm largely contributes to these faults, since developers do not have historical records or examples where to learn from previous code, as it happens in traditional software coding, where software reuse and coding by imitation are reference practices to help in coding better quality software. 
Another issue is the lack of reference measures, such as quality, complexity or coupling metrics, which are extensively used in out-of-chain software production to keep software projects under control \cite{Ibba_2018}. 

The situation is slowly changing for what concerns historical records (even if history is quite recent) of software code, since the Ethereum \bc can now count on up to 1.5 million deployed \scs, which have been used and running in the last few years. Access to the source code of this body of \scs is still a challenge since the transparency and the open access granted by the public blockchains regards only data registered in the blocks, where only contracts' byte-code is available. 

To access the \scs' source code, the developers must resort to other 
means or to code repositories, such as the classical GitHub or similar resources. 
Fortunately, in the last few years, \ES \footnote{https://etherscan.io/} and other web sites have started providing \scs checking as a service, so that Ethereum developers 
can submit their source code to be analyzed and the source code is made available afterword by the website.
However, there is an odd side of the medal for many reasons: the access to this body of knowledge is far from easy and far from fast; it is not structured and organized;
the \scs metrics are not available and must be computed separately~\cite{pierro2019}. All these tasks can and need to be automatized, to save developers' time and work, as well as computational resources. 
Indeed, in the last few years, a number of research papers have been published reporting findings based on \scs' source code, mined from GitHub or some Ethereum block explorer such as \ES~\cite{Mense2018, Amani2018, Tran19, tonelli2018smart}. However, when conducting this kind of empirical research on \scs with data from Ethereum blockchain, the tasks above mentioned need to be carried out by the developers themselves.
The first task is the downloading of the \scs' source code to be analyzed.
One way to download \scs' source code data is to inspect open-source software (OSS) project repositories such as GitHub~\cite{Jaccheri07, pierro2020gasPrice}. 

Another way to perform this task is to use some Ethereum block explorer. 
These web services allow users to find the desired information by directly accessing the Ethereum blocks, by using a unique identifier or sequentially searching several blocks~\cite{Braga2018}.
Some of these Ethereum block explorers provide RESTful Web services, 
which allow the users to obtain a JSON format payload containing various data. 
These data may be related to a current or past state of the Ethereum blockchain: an example may be the list of transaction addresses included in a given block of the Ethereum blockchain.
This activity might be tedious and time-consuming~\cite{Zhou2005}, when conducted by a single user/developer/researcher.
Furthermore, the obtained \scs' data-set can consist of duplicated smart contracts, i.e. \scs having different addresses but with the same code. 

In this work we tackle these problems and propose an organized, easy to use, large and available software repository for Ethereum \scs source code and metrics where users, researchers, \bc start-ups and developers can take advantage of the body of knowledge collected during the last few years. 
This paper thus proposes \SC, a repository that provides the users with an interface, which allows to search and download smart contracts' source code. The user interface is available at the following online address: \SChtmlGUI.
%  a RESTful web service available at this URL: \href{https://smac.ga/}{https://smac.ga/}\ap{too expensive to maintain this service on Amazon, I have to think of an alternative solution such as GraphQL}
The main challenge of the implementation lies in the fact that the Ethereum \bc stores a massive amount of heterogeneous data, \scs included, which enormously grow in time.
For this reason, \SC was designed to be scalable, by adopting the latest cutting-edge technology, such as document-oriented database, graph query language and serverless computing platform~\cite{Kratzke_2020}.

\section{Related Work}\label{sec:related-work}

\subsection{Previous Literature on Software Corpus Analysis}
Gabel and Su~\cite{Gabel2010} built and studied a corpus of open source software written in three of the most widely used languages: C, C++, and Java. 
The corpus contains six thousand software projects corresponding to 430 million lines of source code. The authors measured the degree to which each project of the corpus can be \dquotes{assembled} solely from portions of the corpus, thus providing a precise measure of \dquotes{uniqueness}. Their primary contribution is to provide a quantitative answer to the following question: \textit{how unique is software?}
Our work also aims to answer this question, because many smart contracts written in the Solidity language have code that is a replication of other smart contracts, although presenting different addresses, as we will see in Section~\ref{sec:cleaning-data}.
Our goal is therefore to answer the question: \textit{how unique are smart contracts written in Solidity?} in order to provide a corpus that is composed by smart contracts which can be distinguished from each other.

Tempero and co-authors~\cite{05693210} presented the "Qualitas Corpus", a curated collection of open-source Java systems. The corpus reduced 
the time needed to find, collect, and organize the necessary source code sets to the time needed to download the corpus. The metadata provided with the corpus explicitly indicates the metrics calculated to identify the main features of the source code: the number of code lines, the number of classes, etc.
Our work also aims to present a curated collection of \scs equipped with a set of metadata with the aim of allowing experts in the blockchain field to perform static analysis.

%Allamanis et Sutton~\cite{githubCorpus2013} present a giga-token corpus of Java code from a wide variety of domains. 
%The corpus is made of eleven thousand projects corresponding to 264 million lines of source code.
%By examining the Java corpus, they observe the Zipf's law, usually found in natural languages. 
%Zipf's law states that given some corpus of natural language utterances, the frequency of any word is inversely proportional to its rank in the frequency table.
% head -20 <file.sol> | tr '[A-Z]' '[a-z]' | tr -C '[a-z]' '\n' | sort | uniq -c  | sort -rn|less
% Section~\ref{sec:analyzing-data} investigates this property to see if this law also applies to the smart contracts' corpus analysed in this paper.

% Unlike previous works that concern software corpus formed by projects that do not concern the blockchain, 

\subsection{Static Analysis on Smart Contract Code}
There is a number of scientific publications having as objective the analysis of the \scs' source code, also testifying the scientific community's interest in advancing the knowledge about the characteristics of \scs' code structure.

Hegedus~\cite{Hegedus18} developed a metric calculator for Solidity code, inspired by the work by Tonelli and collaborators~\cite{tonelli2018smart}. The metric calculator uses a parser to generate an abstract syntax tree (AST), on which it computes various software metrics, such as the number of code lines for each \sco, the Cyclomatic Complexity, the number of functions, the number of parameters for each function. This command-line tool is written in Java and is available on GitHub without license indication since February 2018~\footnote{https://github.com/chicxurug/SolMet-Solidity-parser}. By using this tool, he calculated and published software metrics results for 10,206 Solidity \scs source code files written in Solidity languages.
Our work also aims to calculate a set of metadata on the \scs corpus by using a similar software.

Pinna and colleagues~\cite{Pinna2019AMA} performed a static analysis on 10,174 \scs, deployed in the Ethereum blockchain.
The authors showed that some metrics related to smart contracts, such as the number of transactions and the balances follow power-law distributions. 
Also, they reported that software code metrics in Solidity have (on average) lower values but higher variance than metrics values in other programming languages for standard softwares. 
Our work is inspired by their research as \SC is characterized by (some of) the metrics they defined, as we will explain in Section~\ref{sec:modelling-data}.
% Section~\ref{sec:analyzing-data} tries to confirm the conclusion found by their study on the power laws distribution.
%In addition to their work, the paper presents a tool that aims to have a smart contracts' corpus which is continuously growing.
%Furthermore, the research make the smart contracts' corpus  accessible to the research community in two ways: 1) through API calls 2) through a web interface.

Pierro and Tonelli~\cite{PASO} pointed out that even the most experienced users, as software developers of smart contracts are, need to be helped to analyze smart contracts and write a more reliable and secure code. For this reason, an open-source platform~\footnote{https://aphd.github.io/paso/}, called PASO, was proposed as an aid for experts in \scs' static analysis. Their work focused on \Eth blockchain and \scs written in \Sol. The platform PASO facilitates the debugging of \scs by providing software metrics commonly used to comply with coding guidelines.

\subsection{Related Projects}
Other projects, similar to \SC, have been previously developed, to online access \scs' code deployed in the Ethereum blockchain platform.
The projects present specific features and limitations which are summarized in
Table~\ref{tab:related-work}.

\begin{table}[H]
    \caption{Projects' list, main features and limitations}
    \centering
    \label{tab:related-work}
    \def\arraystretch{1.2}
    \begin{tabular}{c|l|l|p{45mm}} % <-- Changed to S here.
      \textbf{Project's name} & \textbf{Home Page} & \textbf{REST API URL} & \textbf{Limitations} \\
      \hline
      GitHub   
      & \href{https://github.com/}{https://github.com/}  
      & \href{https://docs.github.com/en/free-pro-team@latest/graphql}{https://developer.git...}                                                           
      & Some repositories have restricted access.
      \\
      Ethplorer               
      & \href{https://ethplorer.io/}{https://ethplorer.io/}             
      & \href{https://api.ethplorer.io/getAddressInfo/0xff71cb760666ab06aa73f34995b42dd4b85ea07b?apiKey=freekey}{https://api.ethplorer...} 
      &  Requests are limited to 3000/week.    
      \\
      \ES               
      & \href{https://etherscan.io/}{https://etherscan.io/}             
      & \href{https://etherscan.io/apis}{https://etherscan...}  
      & \Scs' addresses are not immediately available.
      \\
      EtherChain
      & \href{https://www.etherchain.org/}{https://www.etherch...}
      & \href{https://www.etherchain.org/api/gasPriceOracle}{https://www.etherch...}
      & \Scs' source code is not available.
      \\
      BlockScout
      & \href{https://blockscout.com/}{https://blockscout.com/}
      & \href{https://blockscout.com/eth/mainnet/api\_docs}{https://blockscout.com...} 
      & \Scs' source code is not available.
      \\
    \end{tabular}
\end{table}

\subsubsection{GitHub}\label{sec:github}
GitHub is the largest collaborative source code hosting site built on top of the Git version control system~\cite{loeliger2012version}. 
The availability of a comprehensive API has made GitHub a target for many software engineering and online collaboration research efforts~\cite{Gousios2017}.
GitHub offers just open-source software to the community. 
In GitHub, there are many works regarding projects written in different programming languages, such as Java, Python and Solidity, which is by far the most commonly used  language to write smart-contracts.

Some GitHub repository collects smart contracts' source code but such code typically has not a direct reference to smart contracts deployed on the Ethereum blockchain and to a corresponding Ethereum address, and it is had to find out if it has been really tested or used on the blockchain. 
Many smart contracts available on GitHub are there just for testing purpose and not for production and 
in many occasions are only deployed on a test net.
Thus for the purpose of the paper, we collect only smart contracts that are deployed in the main network of the Ethereum blockchain. 
Besides, GitHub lacks some information needed to users to select or filter the smart-contracts, and the researcher might not want to select the smart-contracts' source code randomly or on the basis of their forks or likes number. 
It is highly probable that the users, especially if developers or researchers, want to access smart-contracts' source code having the possibility to choose on the basis of specific software metrics and on its real usage on the blockchain. 
%This, for sure, is a general idea that is not valid just for the smart-contracts' code but also for any software project.

\subsubsection{Ethereum Block Explorers}
\label{sec:ethereum-block-explorers}
Ethereum block explorers are platforms that allow the users to explore and search the Ethereum blockchain for transactions, addresses, tokens and other activities taking place on the Ethereum blockchain~\cite{Bistarelli20}.
Unlike GitHub, the Ethereum block explorers allow accessing only Ethereum data used in the Ethereum blockchain and thus smart contracts' real use-cases. 
To date, in the market, there are different Ethereum block explorers, as for instance:

\begin{itemize}
    \item \textbf{Ethplorer}\footnote{https://ethplorer.io/} provides an API to access many Ethereum data, such as the balances for specified token, the description of a specific address but it does not allow to access to the smart contracts' code.
    The full documentation of the Ethpoler API is available at the following address~\footnote{https://github.com/EverexIO/Ethplorer/wiki/Ethplorer-API}.
    The Requests to API are limited to 5 per second, 50/min, 200/hour, 2000/24hours, 3000/week.
    \item \textbf{EtherChain}\footnote{https://etherchain.org/} is an explorer for the Ethereum blockchain.
    Unlike Ethplorer, it claims to provide smart contract code, even though it actually displays the contract byte-code and the Constructor arguments, for a specific smart contract's address.
    EtherChain provides API just to access the Oracle gas price predictions~\footnote{https://www.etherchain.org/api/gasPriceOracle}, but not the Ethreum data. 
    If the users want to gather Ethereum data from EtherChain, they need to parser the HTML code.
    \item \textbf{BlockScout}\footnote{https://blockscout.com/poa/xdai/} provides an API to access the Ethereum data. 
    It claims to have an API to access only the source code of few verified smart contracts. Anyway the addresses list of the verified smart contracts is not available in BlockScout.
    %and when the Author of the paper tried to download it using the API to get the source code of the smart contract they get a json object with no valuable information about the source code.
    \item \textbf{\ES} allows to explore and search the Ethereum blockchain for smart-contracts. However, when downloading the smart-contracts' source code, the block explorer presents some limitation. First, smart-contracts' data and number are huge (on the Giga scale, based on our estimation), but there is a limited API rate of 100 submissions per day per user to retrieve just a smart contract, making the complete download of the data an impossible endeavour~\footnote{https://etherscan.io/apis\#contracts}. Second, the \ES's API does not provide facilities to obtain a list of the smart-contracts's address, as the existing API calls mainly allow navigation from one block to another. Third, a researcher cannot directly and easily explore the smart-contract's source code, but rather has to first inspect any block in Ethereum and then look for all the transactions that involve an address associated with the smart-contract.
\end{itemize}

\section{Research Methodology}
\begin{figure}[H]
    \centering
    {\includegraphics[width=0.9\linewidth]{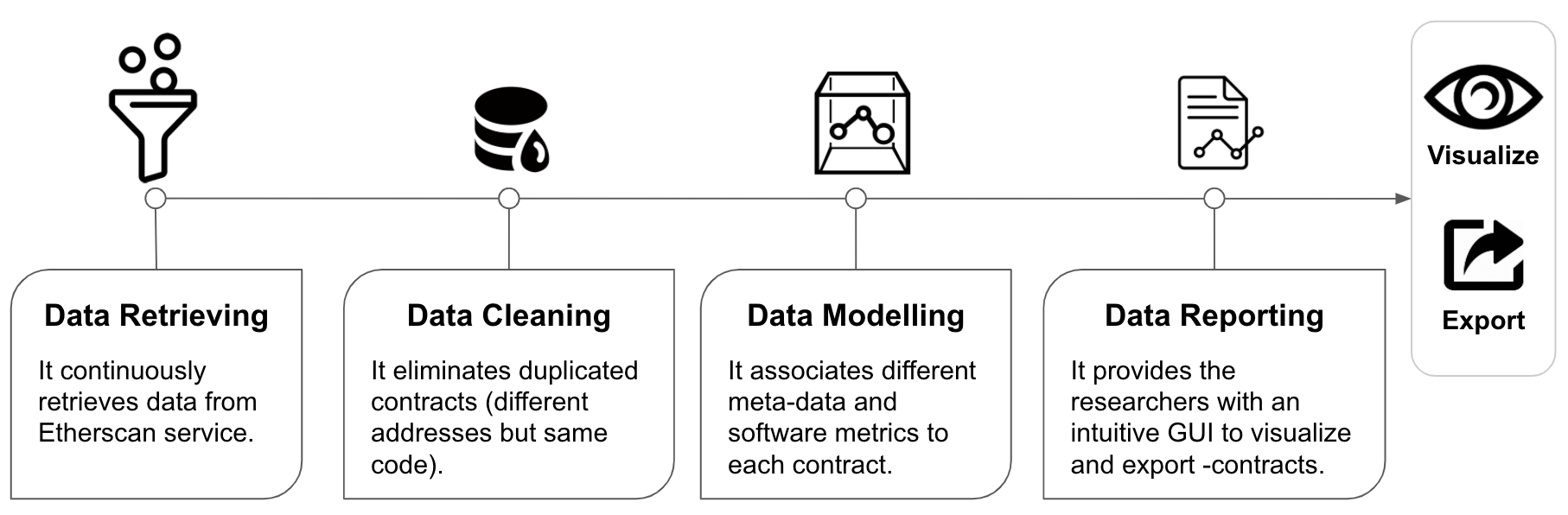}}
    \caption{Smac-Corpus's pipeline model}
    \label{fig:pipeline-model}
\end{figure}

\SC has been designed to provide the users with a reasoned repository, i.e. a repository which is not just a web space where to collect them but also and mainly a service to help the researchers to filter and analyze the Smart contracts' source code. To this aim, \SC has been planned to perform four main automatic operations on Smart contracts' source code (data):
\begin{enumerate*}
    \item Data Retrieving,
    \item Data Cleaning,
    \item Data Modelling,
    \item Data Querying.
\end{enumerate*}
Figure~\ref{fig:pipeline-model} shows the \SC's pipeline of the operations.

\subsection{Retrieving Data}
We collected smart-contracts' source code, smart-contracts' application binary interface (ABI) and smart-contracts' byte-code through the \ES web site, which makes available the source code of a subset of verified \scs 
deployed on the Etehreum blockchain, though in a laborious way. We instead made this task easier and automatic via a retrieving data script available at the following address~\footnote{https://github.com/aphd/solidity-metrics/tree/master/examples}. 
During this phase the blockchain blocks are automatically inspected.
Each block is formed by a list of transactions between two different blockchain addresses, which can refer to a wallet or to a smart-contract.
The script looks for addresses that refer to a smart-contract and when the source code is available, it downloads the smart-contract's source code, the ABI and the byte-code.
The data coming from the source code are not immediately available as they are embedded in the HTML code of the web page provided by \ES. Therefore, the script removes the HTML tags and stores the code cleaned up.

\begin{figure}[H]
    \begin{subfigure}{.33\textwidth}
        \includegraphics[width=.9\linewidth]{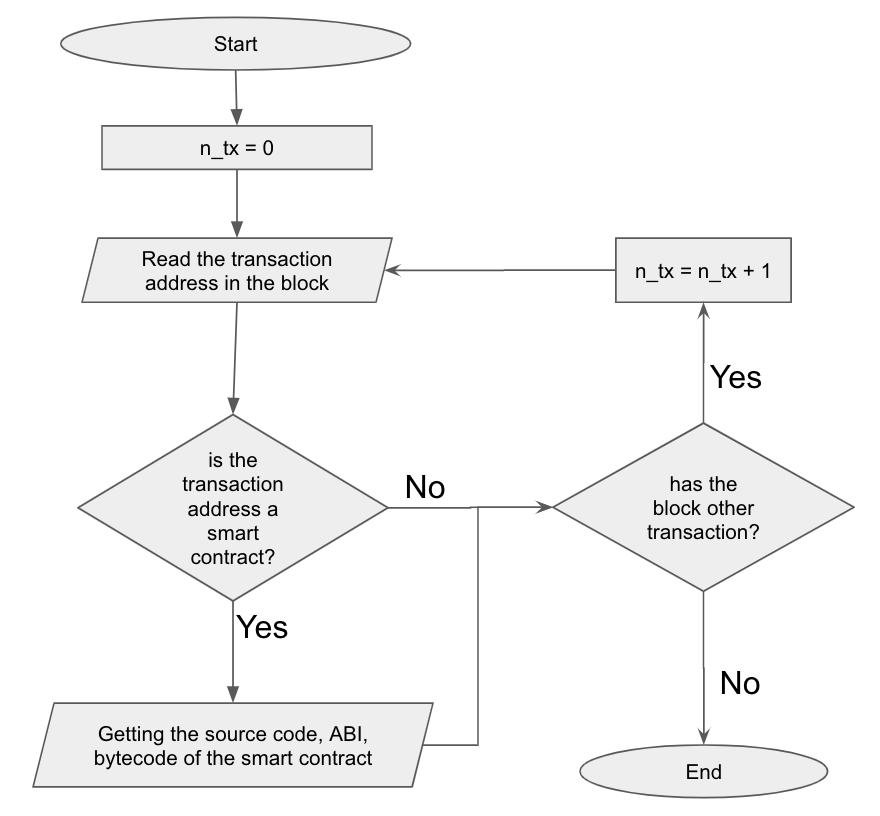}
        \caption{Transactions' list in a block}
        \label{fig:retrieving-data-a}
    \end{subfigure}%
    \begin{subfigure}{.33\textwidth}
        \includegraphics[width=.9\linewidth]{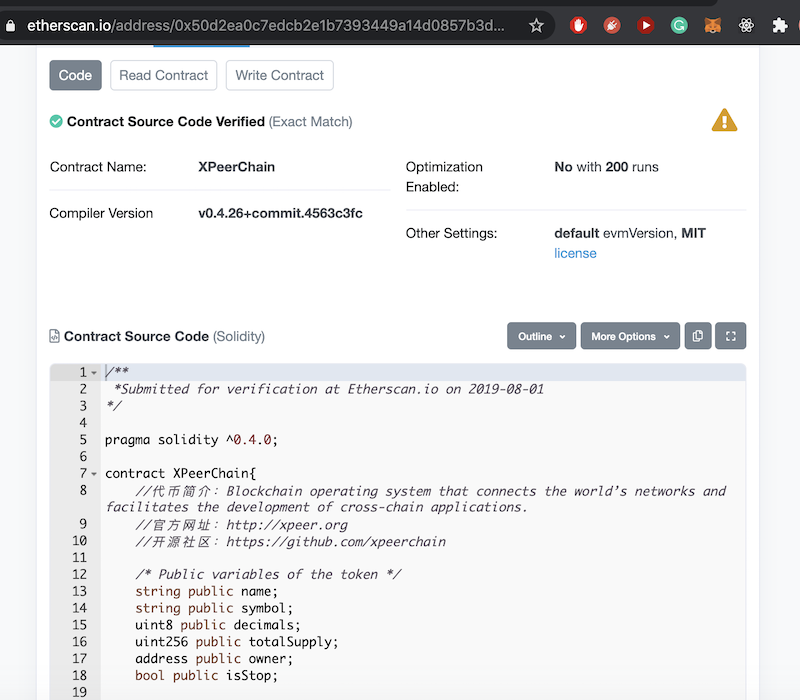}
        \caption{Smart contract's webpage code}
        \label{fig:retrieving-data-b}
        \end{subfigure}
    \begin{subfigure}{.33\textwidth}
        \includegraphics[width=.9\linewidth]{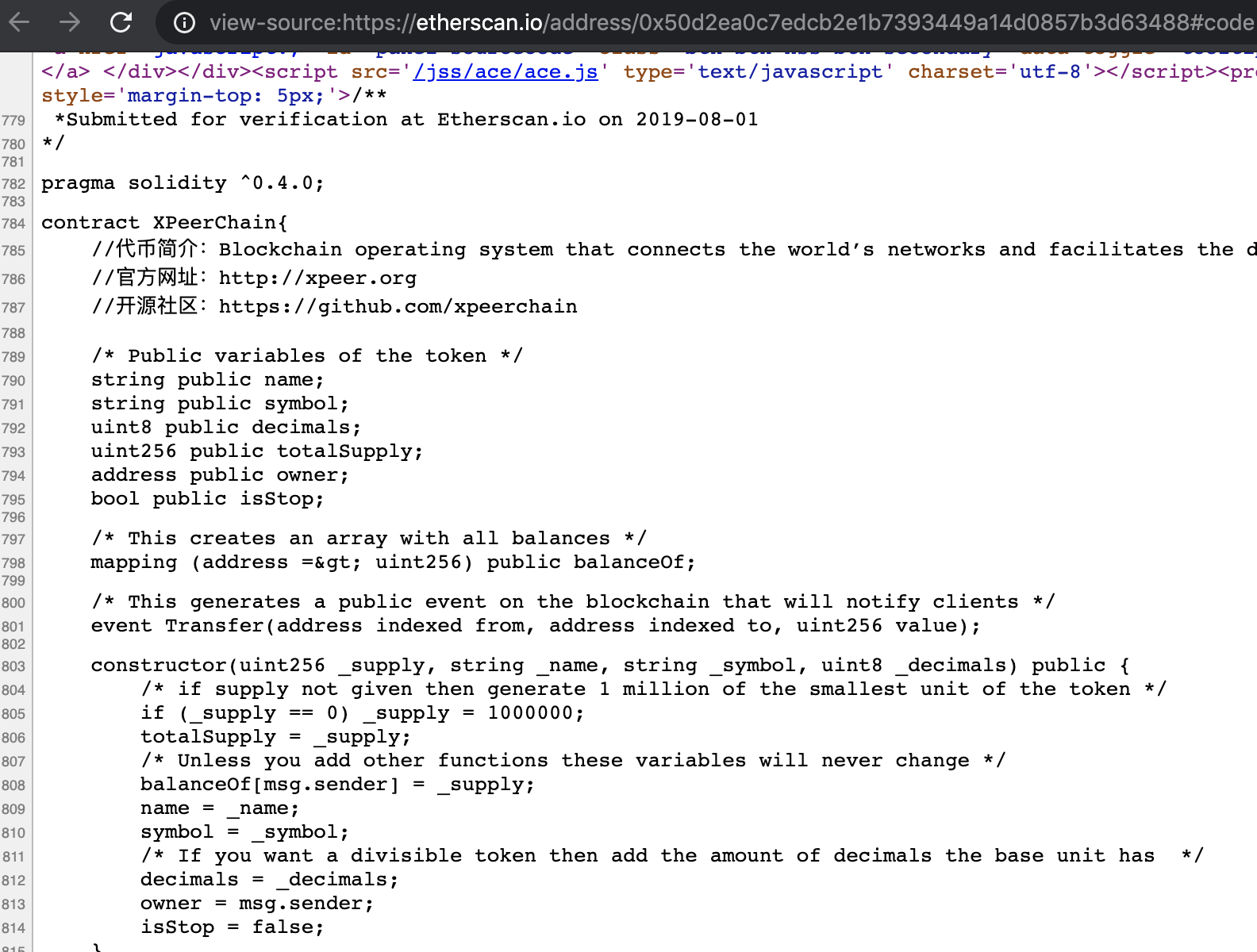}
        \caption{Smart contract's source code}
        \label{fig:retrieving-data-c}
        \end{subfigure}
    \caption{Data Retrieving Pipeline. The figures~\ref{fig:retrieving-data-a},~\ref{fig:retrieving-data-b},~\ref{fig:retrieving-data-c} shows three different phases to retrieve the smart contracts.}
    \label{fig:paso-gui}
\end{figure}

Figure~\ref{fig:retrieving-data-a} shows how \SC finds the smart-contracts' list in a given block. Figure~\ref{fig:retrieving-data-b} shows the HTML page where the smart contract code is available. The HTML page containing the smart contract code and the HTML tags is downloaded. Figure~\ref{fig:retrieving-data-c} shows the HTML code that will be processed to remove the HTML tags and save just the Solidity source code of the \sco.

The smart-contracts' code is stored in the filesystem of the \SC server. 
Due to the quota limits on queries per second (\ES web site allows few connections per second), \SC contains only a portion of all available smart contracts.
However, the retrieving data phase is continuously collecting data, starting from December 10, 2019. 
To date, thirty thousand of smart-contracts (source code, ABI and byte-code) have been downloaded and made available through \SC.

\subsection{Cleaning Data}
\label{sec:cleaning-data}
Each smart-contract in the Ethereum blockchain is distinguished from any other smart-contract as it is identified by a unique address, i.e. a hash of 160 bits, and its byte-code is stored on the blockchain~\cite{Chi91}. 
Indeed, each time a smart-contract is deployed in the network, either in the main or in the test network, a unique address is associated with the smart-contract even in the case the source code of two or more smart-contracts is the same. However, this is a problem for the analysis of the software metrics, because the smart-contracts are distinguished only on the basis of their address and not of their content. Therefore, \SC eliminates double contracts in order to provide a clean smart-contracts' corpus, where to perform the analysis. To this aim, duplicate \scs have been defined on the basis of their content, i.e. having the same code despite presenting different address. 

\subsection{Modelling Data}
\label{sec:modelling-data}
\begin{figure}[H]
    \centering
    {\includegraphics[width=0.48\linewidth]{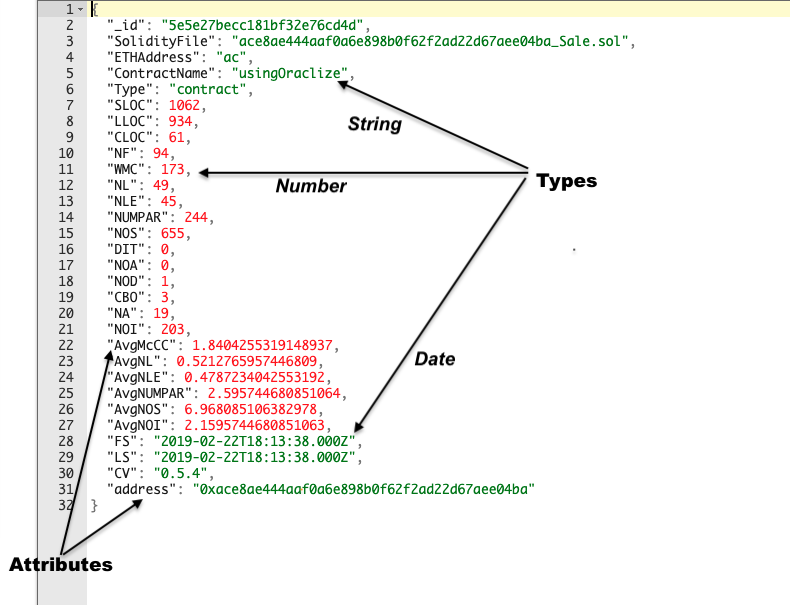}}
    \caption{Smart-Corpus's database schema}\label{fig:smac-nosql-schema}
\end{figure}

Unlike the tools discussed in the related work Section~\ref{sec:related-work}, \SC associates different metrics (intrinsic metrics and extrinsic metrics) to the smart-contracts, aiming to facilitate the selection of a smart-contracts' set that meets precise requirements. The metrics associated with the smart-contracts are then stored in a document-oriented database.
Figure~\ref{fig:smac-nosql-schema} shows the database schema of a smart contract.

% The meta-data of the smart-contract, as discussed in~\cite{} are devised in two categories: intrinsic cues and extrinsic cues.

\subsubsection{Smart Contracts' Intrinsic Metrics}
\label{sec:intrinsic-metrics}
The smart contracts' intrinsic metrics are \scs' software metrics which depend on internal properties of the smart contracts' code, such as the number of lines of code, modifiers, payable, etc.
Table~\ref{tab:intrinsic-cues} shows the \scs' intrinsic software metrics.

\begin{table}[H]
        \caption{Smart-contracts' intrinsic metrics}
        \label{tab:intrinsic-cues}
        \centering
        \def\arraystretch{1.3}
        \begin{tabular}{l|p{105mm}}
            \textbf{Name} & \textbf{Description} \\
            \hline
            Pragma & \dquotes{Pragma} indicates version indicates which version of Solidity compiler is used to prevent issues with future compiler versions.\\
            SLOC & \dquotes{SLOC} indicates the number of lines in a \scs' source code. \\
            Modifiers & \dquotes{Modifiers} indicates the number of function modifiers in a \sco. \\
            Payable & \dquotes{Payable} indicates the number of payable functions in a \sco. \\
            Mapping & \dquotes{Mapping} indicates the number of variables of mapping types in a \sco. \\
            Address & \dquotes{Address} indicates the number of variables of address types in a \sco. \\
        \end{tabular}
\end{table}

\subsubsection{Smart Contract Extrinsic Metric}
\label{sec:extrinsic-metrics}
The \scs' extrinsic metrics are properties depending on external factors rather than the code itself, such as the number of transactions executed to the smart-contracts or the number of tokens associated with the smart contracts.
Table~\ref{tab:extrinsic-cues} shows the smart-contracts' extrinsic metrics.
\begin{table}[H]
        \caption{Smart-contracts' extrinsic metrics}
        \label{tab:extrinsic-cues}
        \centering
        \def\arraystretch{1.3}
        \begin{tabular}{l|p{105mm}}
            \textbf{Name} & \textbf{Description} \\
            \hline
            Transactions & \dquotes{Transactions} represents the total number of transactions generated by the smart contract (sent or received). \\
            Balance & \dquotes{Balance} is the amount of crypto coins associated with a smart-contract address. \\
            EtherValue & \dquotes{EtherValue} is the dollar value associated with a smart-contract address.\\
            Token & \dquotes{Token} is the value for each token associated with a smart-contract address. \\
            Last\_seen & \dquotes{Last\_seen} is the timestamp of the last time the smart-contract was used (sent or received).\\
            First\_seen & \dquotes{First\_seen} is the timestamp of the first time the smart-contract was used (sent or received).\\
        \end{tabular}
\end{table}

\subsection{Filtered Data}
\begin{figure}[H]
    \centering
    {\includegraphics[width=0.78\linewidth]{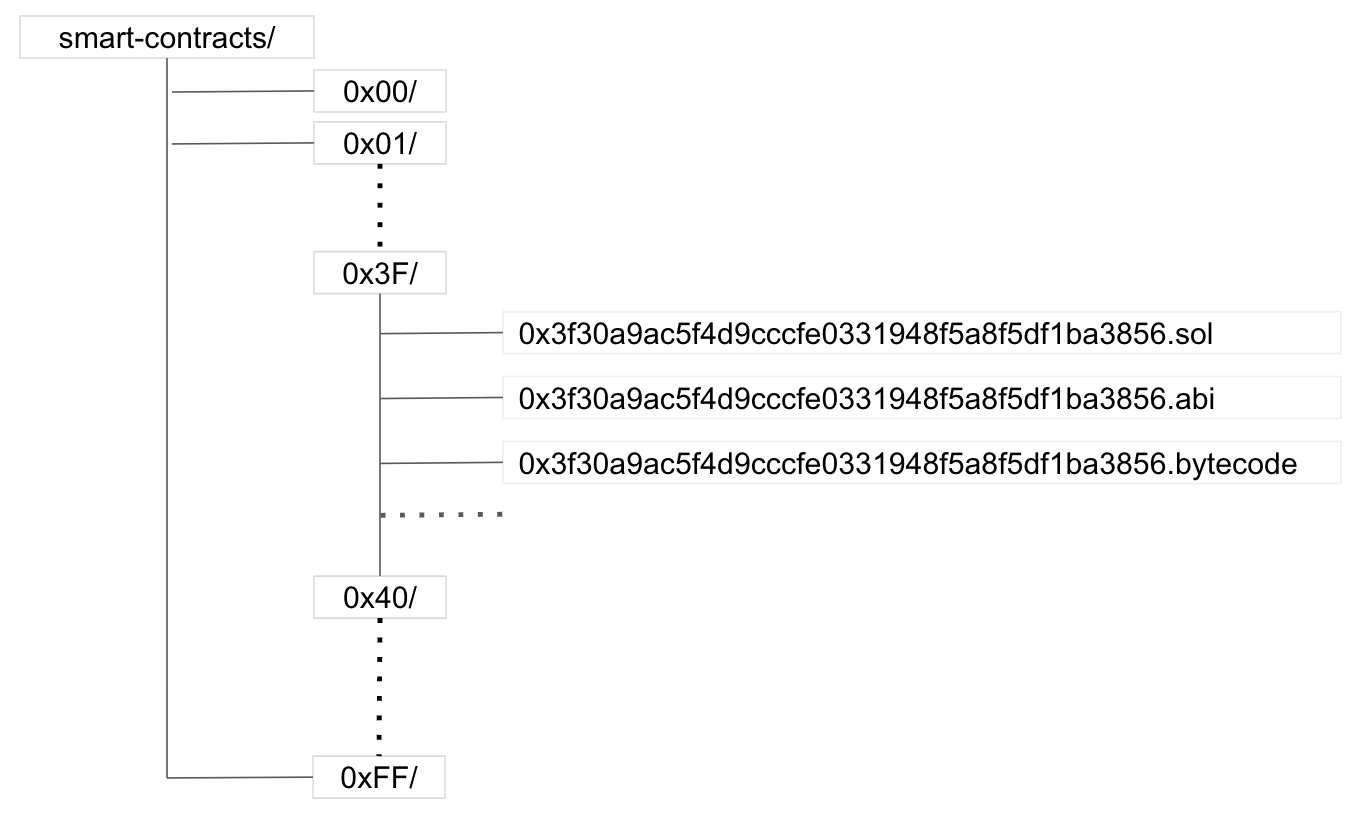}}
    \caption{Smart-contracts' directory structure.}\label{fig:directory-structure}
\end{figure}

The smart contracts' source code is stored in a file system and is organized in folders and subfolders to ease the navigation.
Figure~\ref{fig:directory-structure} shows the sub-directory structure.
The first leaf corresponds to the first two letters of the smart contract address and then each directory contains the file having as name the full address of the smart contract and three different extensions, respectively .sol for the \Sol source code, .abi for the ABI, and .bytecode for the byte code.

The metadata (both the intrinsic and extrinsic metrics) are stored in a document-oriented database: MongoDB~\cite{Chodorow2013}. 
%MongoDB-The-Definitive-Guide.pdf
The choice to use a Document-oriented database instead of a relational database such as Mysql is based on:
\begin{itemize}
    \item Relational databases are prone to deterioration when data-sets overcome a size threshold, while a document-oriented database such as MongoDB comes with inbuilt load balancer, which makes it a better solution in applications with high data load~\cite{Diogo_2019}. We update MongoDB each day to generate the data archive. 
    \item Unlike relational databases where data is stored in rows and columns, document oriented databases store data in documents. The documents typically use a structure similar to JSON (JavaScript Object Notation), they indeed provide a natural way to model data that is closely aligned with object oriented programming. Each document is considered as an object in object oriented programming, similarly each document is a JSON in document-oriented database. The concept of schema in document databases is dynamic: every document might contain a different number of fields. This is useful when modeling unstructured and polymorphic data. Also, document databases allow robust queries: any combination of fields in the document can be combined for querying data~\cite{Baker_Effendi_2020}. 
\end{itemize}

\subsection{User Interface}
The \SC' graphical user interface (GUI) allows users to access the \scs' repository.
There are two way to access the \scs' repository: through the \dquotes{HTML user interface} and through a \dquotes{GraphiQL application}, both of them via a Web browser.

\subsubsection{Smart Corpus HTML user interface}
\label{sec:html-gui}
\begin{figure}[H]
    \centering
    {\includegraphics[width=0.88\linewidth]{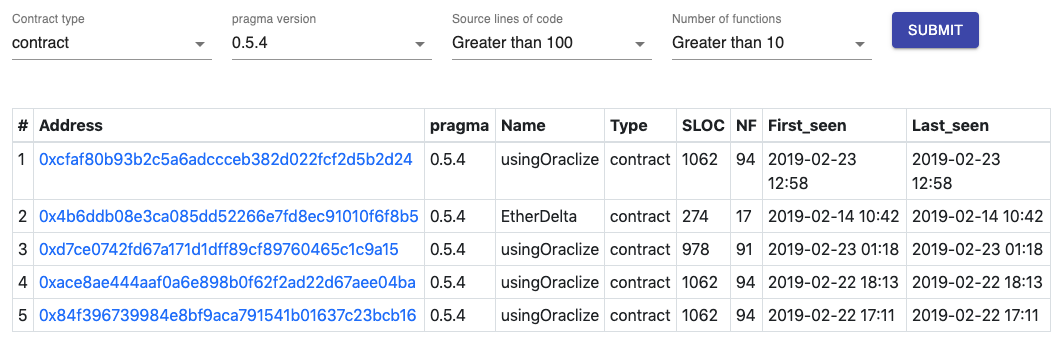}}
    \caption{\SC's User Interface}
    \label{fig:html-gui}
\end{figure}

The \SC HTML user interface is publicly available since January 2020~\footnote{\SChtmlGUI}.
Figure~\ref{fig:html-gui} shows the different components of the GUI.
\begin{itemize}
    \item At the top, the user can find the form where to filter the \scs. 
    The form is made of a number of drop-down lists, each one corresponding to a different metric and a submit button to perform the research.
    The GUI form allows the user to inspect smart contracts based on some metadata, such as the \dquotes{pragma version}, and software metrics, such as the numbers of \dquotes{modifiers} and/or the numbers of \dquotes{payable}.
    \item Below the form, the smart contracts filtered by the user are displayed. For readability reasons, only a part of the smart-contracts metrics are presented in the table layout format. Each column header in the table indicates the name of a metric associated to smart-contracts. While the HTML-GUI displays just some metrics, the user can access to all the metrics and to the smart-contracts' source code by selecting the checkbox displayed on the right of the smart-contract address and clicking on the red button "download". The user can also access the original repository where the smart-contract was retrieved, i.e. the \ES service.
\end{itemize}

\subsubsection{\SC GraphQL application}
\label{sec:graphql-gui}
\begin{figure}[H]
    \centering
    {\includegraphics[width=0.9\linewidth]{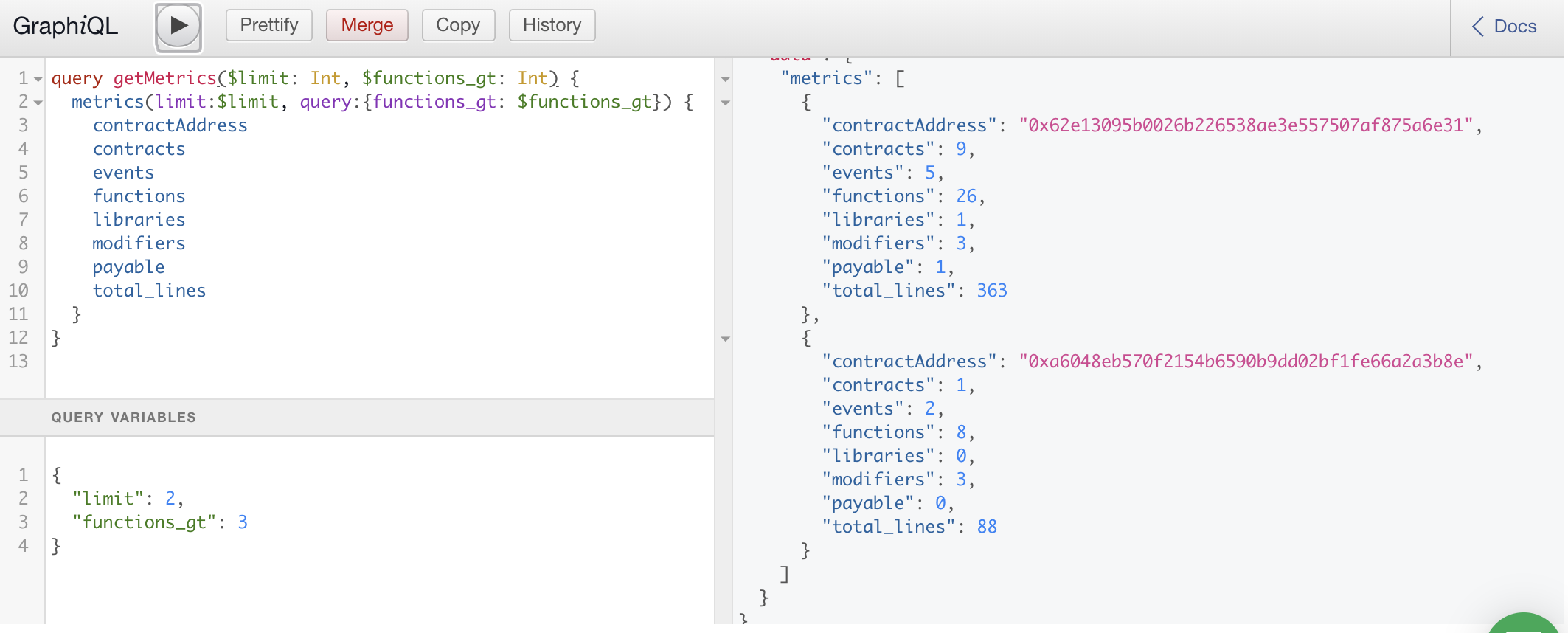}}
    \caption{Example with the use of variables to filter a query result with GraphQL.}
    \label{fig:graphql}
\end{figure}

Graph Query Language (GQL) is a full data query language to implement web-based services, centered on high-level abstractions, such as schemas, types, queries, and mutations. GQL is a domain-specific language internally developed in Facebook from 2012 onward and publicly announced in 2015, with the release of a draft language specification.
The language was conceived with the following goals:
\begin{itemize}
    \item To reduce possible overload of data transfer relative to REST-like web service models, in terms of both the amount of data unnecessarily transferred, and the number of separate queries required to do it.
    \item To reduce the potential of errors caused by invalid queries on the part of the client. In particular, with the GQL application, the user can execute \dquotes{type introspection}, i.e. the user can examine the type or properties of an object at runtime. For example, thanks to the introspection queries the user can find out both the intrinsic and the extrinsic metrics associated with a specific \sco while typing the query.
\end{itemize} 
Figure~\ref{fig:graphql} shows an example query and its result.

\SC GQL application, unlike the \SC HTML user interface
%, is not yet publicly available, because it 
is still in development and testing phase. 
However, the \SC GQL application source code is publicly available and can be downloaded and deployed on any platform having the software requirements specified in its documentation~\footnote{https://github.com/aphd/smac-corpus-api}. 

\subsection{Use Case}
A use case for \SC might concern a researcher interested in the static analysis of smart contracts. 
For example, the researcher might be interested in performing an analysis of smart-contracts written with a particular version of the Solidity language, 6.0, and having at least a payable function in the \sco.
The research of smart contracts that meets these requirements would be very expensive in terms of time, work and computational resources, using a service like \ES. 
Instead, thanks to the \SC, the user needs to perform only a few steps, as described below:
\begin{itemize}
    \item connect to the service through the link: \SChtmlGUI
    \item select the option "version 6.0" from the drop-down menu entitled "pragma version".
    \item select the option "greater than zero" from the drop-down menu entitled "number of payables".
    \item submit the form by clicking on the button "submit".
\end{itemize}

After few seconds, depending on the number of smart-contracts that meet the requirements specified by the user, the smart-contracts' addresses and the metrics values will be displayed in a table layout format ready to be downloaded.

\section{Results}

% This section may be divided by subheadings. It should provide a concise and precise description of the experimental results, their interpretation as well as the experimental conclusions that can be drawn.

\begin{figure}[H]
    \centering
    {\includegraphics[width=0.88\linewidth]{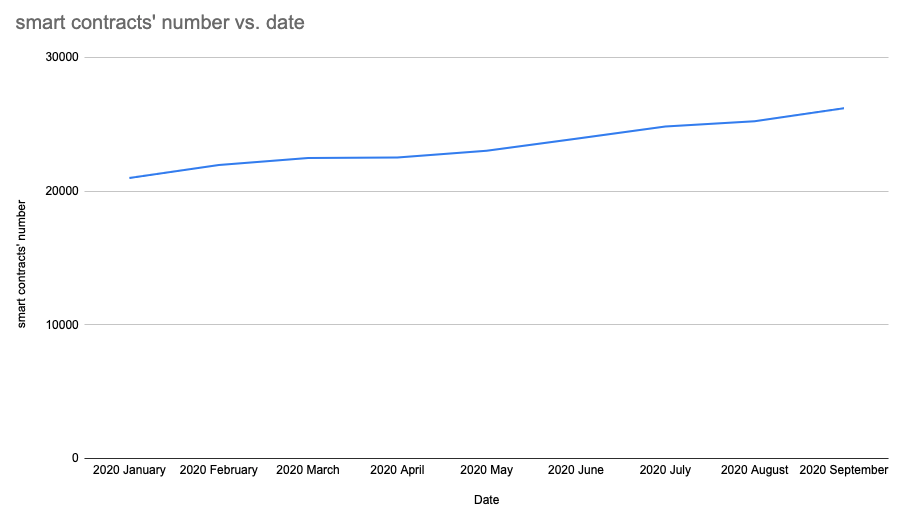}}
    \caption{Number of smart-contracts' collected in \SC.}
    \label{fig:trend-plot}
\end{figure}

The \SC has been in use for 10 months, since December 2019 and 100K smart contracts have been downloaded via the user interface. 
Until the paper was written (October 2020), \SC is a curated corpus of 30K \scs's source codes, ABI and byte-codes with related metadata and software metrics.
As the time passes, the \SC is continuously increasing at a rate of 100 smart contracts per day.
Figure~\ref{fig:trend-plot} shows the number of smart-contracts' source codes, ABI and byte-codes retrieved per day since the \SC was deployed for the first time.
For each smart contract \SC computed extrinsic and intrinsic metrics, as described in Sections~\ref{sec:extrinsic-metrics} and~\ref{sec:intrinsic-metrics}.

Summing up, the \SC has two GUIs to access data, the HTML GUI and the GraphQL interface.
The HTML GUI is described in Section~\ref{sec:html-gui}, while the GraphQL interface is described in Section~\ref{sec:graphql-gui}.
The GraphQL interface gives blockchain researchers the ability to request for exactly what they need.
The user can directly access the results via GraphQL interface, as shown in Figure~\ref{fig:graphql}. 

% It is currently used as base for empirical study on software makeanalysis\ap{add reference}

Unlike the existing repositories (see Section~\ref{sec:ethereum-block-explorers}) which makes available the source code in a laborious way, 
\SC instead made this task easier and faster.
Indeed, one of the advantages of using \SC lies in the fact that it can reduce the costs in performing the smart contract static analysis. 
For example, it can be used to easily analyze design and programming patterns for the smart contract programming language.
\SC allows to analyze how the industry companies use the Solidity programming language to solve concrete problems in different application areas, such as Healthcare, Insurance, Transportation, Government, Entertainment and Energy.
%%%%%%%%%%%%%%%%%%%%%%%%%%%%%%%%%%%%%%%%%%
% \section{Discussion}

%%%%%%%%%%%%%%%%%%%%%%%%%%%%%%%%%%%%%%%%%%

\section{Conclusions and Future Works}
In this paper, we described the \SC project, an effort to bring smart contracts data (source codes, ABIs and byte-codes) to the hands of the research community, providing help for reproducible research and a less time-consuming way to gather data and perform static analysis. 
The project has already stored several megabytes of data, which correspond to about thirty thousand of smart contracts.
This work corresponds to 10 months of data retrieving that are made available to the blockchain scientific community and the blockchain developers in a few seconds.
The \SC data-set has strong potential to provide an interesting venue for research in many software engineering areas including, but not limited to, the best practices for \Sol software development, distributed collaboration, and code paternity and attribution.
The \SC project is in its initial stage of development, but it can already provide useful insight for researchers on smart contracts' coding and everyday use in the \bc.
The corpus will continue to be expanded in content and in the provision of intrinsic and extrinsic metrics, this becoming more and more representative of the Solidity code actually used in the \bc community.

%%%%%%%%%%%%%%%%%%%%%%%%%%%%%%%%%%%%%%%%%%
\vspace{6pt} 

%%%%%%%%%%%%%%%%%%%%%%%%%%%%%%%%%%%%%%%%%%
%% optional
%\supplementary{The following are available online at \linksupplementary{s1}, Figure S1: title, Table S1: title, Video S1: title.}


%%%%%%%%%%%%%%%%%%%%%%%%%%%%%%%%%%%%%%%%%%

%%%%%%%%%%%%%%%%%%%%%%%%%%%%%%%%%%%%%%%%%%

%%%%%%%%%%%%%%%%%%%%%%%%%%%%%%%%%%%%%%%%%%

%%%%%%%%%%%%%%%%%%%%%%%%%%%%%%%%%%%%%%%%%%

%%%%%%%%%%%%%%%%%%%%%%%%%%%%%%%%%%%%%%%%%%
%% optional

%%%%%%%%%%%%%%%%%%%%%%%%%%%%%%%%%%%%%%%%%%
%% optional
\appendix
\section{Appendix}
\subsection{Queries}
Listing~\ref{lst:gql-1} displays a GQL query that returns the smart-contracts's address having more than 20 methods defined in a contract. 
Listing~\ref{lst:gql-2} displays the query results in the JSON format. 
The query output, in addition to the smart-contract's address, contains various information (intrinsic metrics) such as the number of events, the number of functions, the number of modifiers and the number of payables, as specified by the query~\ref{lst:gql-1},

\begin{lstlisting}[caption={A GQL query for displaying intrinsic metrics.},label={lst:gql-1},language=SQL]
    {
        metrics(query:{functions_gt: 20}) {
          adress
          events
          functions
          modifiers
          payable
        }
      }
\end{lstlisting}

\begin{lstlisting}[caption={A GQL result displaying intrinsic metrics.},label={lst:gql-2},language=SQL]
    {
        "data": {
          "metrics": [
            {
              "contractAddress": "0xb7f4c286851cbf0cbf2fe8ebf40412b196c0e8ad",
              "events": 7,
              "functions": 27,
              "modifiers": 1,
              "payable": 1
            },
            {
              "contractAddress": "0x755cebe8cc53c7cb1e1bb641026a17d37d4aea91",
              "events": 4,
              "functions": 31,
              "modifiers": 1,
              "payable": 4
            },
            {
              "contractAddress": "0xb92aa4a864daf0d6a509e73a9364feba44384965",
              "events": 3,
              "functions": 24,
              "modifiers": 1,
              "payable": 1
            },

            ...

        }
    }
\end{lstlisting}

Listing~\ref{lst:gql-exstrinsic-metrics-query} displays a GQL query that returns some extrinsic metrics of a specific smart-contracts's address. 
Listing~\ref{lst:gql-exstrinsic-metrics-result} displays the query results in the JSON format. 
The query output, in addition to the smart-contract's address, contains information such as the total number of transactions generated by the smart contract, the amount of crypto coins associated with the smart-contract's address specified by the query~\ref{lst:gql-exstrinsic-metrics-query},

\begin{lstlisting}[caption={A GQL query for displaying exstrinsic metrics.},label={lst:gql-exstrinsic-metrics-query},language=SQL]
    {
        metrics(query:{address_eq: "0x536c7efeebff067a69393133b1c87a163a6b0598"}) {
          adress
          transactions
          balance
        }
      }
\end{lstlisting}

\begin{lstlisting}[caption={A GQL result displaying exstrinsic metrics.},label={lst:gql-exstrinsic-metrics-result},language=SQL]
    {
        "data": {
          "metrics": [
            {
              "contractAddress": "0x536c7efeebff067a69393133b1c87a163a6b0598",
              "transactions": 639 ,
              "balance": 0 Ether
            }
          ]
        }
    }
\end{lstlisting}

%%%%%%%%%%%%%%%%%%%%%%%%%%%%%%%%%%%%%%%%%%

\bibliographystyle{unsrt}  
\bibliography{main}

% The following MDPI journals use author-date citation: Arts, Econometrics, Economies, Genealogy, Humanities, IJFS, JRFM, Laws, Religions, Risks, Social Sciences. For those journals, please follow the formatting guidelines on http://www.mdpi.com/authors/references
% To cite two works by the same author: \cite{ref-journal-1a} (\citeyear{ref-journal-1a}, \citeyear{ref-journal-1b}). This produces: Whittaker (1967, 1975)
% To cite two works by the same author with specific pages: \cite{ref-journal-3a} (\citeyear{ref-journal-3a}, p. 328; \citeyear{ref-journal-3b}, p.475). This produces: Wong (1999, p. 328; 2000, p. 475)

%%%%%%%%%%%%%%%%%%%%%%%%%%%%%%%%%%%%%%%%%%
%% optional
%% for journal Sci
%\reviewreports{\\
%Reviewer 1 comments and authors’ response\\
%Reviewer 2 comments and authors’ response\\
%Reviewer 3 comments and authors’ response
%}

%%%%%%%%%%%%%%%%%%%%%%%%%%%%%%%%%%%%%%%%%%
\end{document}

%% file: macros.tex
\newcommand{\commented}[1]{}

\newcommand{\dquotes}[1]{``#1''}

\usepackage{url}            
\makeatletter
\def\url@leostyle{%
  \@ifundefined{selectfont}{\def\UrlFont{\sf}}{\def\UrlFont{\small\sffamily}}}
\makeatother
\providecommand{\keywords}[1]
{
  \small	
  \textbf{\textit{Keywords---}} #1
}
\usepackage{amssymb,amsfonts}
\usepackage{algorithmic}
\usepackage{textcomp}
\usepackage[utf8]{inputenc}
\usepackage{xspace}
\usepackage{booktabs}
\usepackage{flushend}
\usepackage{natbib}
\setcitestyle{numbers}
\usepackage{subcaption}
\usepackage[inline]{enumitem}
\usepackage{listings}
\usepackage{float}
\usepackage{hyperref}
\usepackage{graphicx}